\documentclass{ledger}

\newcommand{\thefirstpagenum}[0]{X}
\newcommand{\thelastpagenum}[0]{X}
\newcommand{\theyear}[0]{20XX}
\newcommand{\thevol}[0]{X}
\newcommand{\thedoi}[0]{DOI 10.5195/LEDGER.\theyear.XXX}
\newcommand{\ledgerpages}[0]{\thefirstpagenum-\thelastpagenum}

\hypersetup{pdfauthor={Henriques Pereira, Marco; 
Hafner, Matthias; Dietl, Helmut; Beccuti, Juan}, pdftitle={The four types of stablecoins:
A comparative analysis
}}

\title{The four types of stablecoins:\\
A comparative analysis}
\author{Matthias Hafner\thanks{M. Hafner (matthias.hafner@swiss-economics.ch) is Managing Economist at Swiss Economics, Switzerland.}, Marco Henriques Pereira\thanks{Corresponding Author},\thanks{M. Henriques Pereira (marco.pereira@business.uzh.ch) is PhD candidate and researcher at the University of Zurich, Switzerland.} Helmut Dietl\thanks{H. Dietl (helmut.dietl@business.uzh.ch) is Professor for Services \& Operations Management at the University of Zurich, Switzerland.}, Juan Beccuti\thanks{J. Beccuti (juanbeccuti@gmail.com) is Economics Researcher at Informal Systems, Canada.}}

\pretitle{
  \centering \selectfont FULL PAPER, ChainScience 2023 \par
  \fontsize{24pt}{28pt}\selectfont} 

\begin{document}

\maketitle

\thispagestyle{pagefirst}

\begin{abstract}
Stablecoins have gained significant popularity recently, with their market cap rising to over \$180 billion. However, recent events have raised concerns about their stability. In this paper, we classify stablecoins into four types based on the source and management of collateral and investigate the stability of each type under different conditions. We highlight each type’s potential instabilities and underlying tradeoffs using agent-based simulations. The results emphasize the importance of carefully evaluating the origin of a stablecoin’s collateral and its collateral management mechanism to ensure stability and minimize risks. Enhanced understanding of stablecoins should be informative to regulators, policymakers, and investors alike.

\end{abstract}

\section{Introduction}
\label{sec:1}
Stablecoins, a type of cryptocurrency designed to maintain a stable value relative to a particular asset or group of assets, have experienced tremendous growth in recent years \cite{bullmann2019search, kolodziejczyk_stablecoinstable_2020}. The increasing demand for stablecoins has led to a surge in the number of stablecoins available, and the market cap for stablecoins has risen to over \$180 billion in just a few years.\cite{duan_instability_2023}.

However, recent events such as the crash of TerraUSD and the de-pegging of USDC have raised questions about the stability of stablecoins. As a result, there is a growing need for a better understanding of how stable these coins actually are \cite{chohan_are_2019} and the differences between various types of stablecoins. Furthermore, the rapid growth and questions about stability have raised concerns among regulators, and there is currently ongoing discussion about how to regulate stablecoins \cite{kaloudis_draft_2023}. As there is still limited knowledge about stablecoins, this has become a crucial area of research.

This paper proposes a novel categorization of stablecoins and investigates the stability of each type under different conditions. Our categorization is based on two key dimensions, resulting in a 2 × 2 matrix that classifies stablecoins into four types. The first dimension pertains to the \textit{source} of collateral, i.e., whether it comes from an exogenous source (such as fiat money and gold reserves) or an endogenous source (such as crypto assets like sUSD and TerraUSD). The second dimension pertains to collateral \textit{management}, i.e., whether a central entity or mechanism manages pooled collateral and decides when to expand or contract supply or whether individuals manage their collateral decentrally and mint and burn stablecoins to adjust supply. This categorization provides an economical and comprehensive way to classify stablecoins.

To assess the stability of each stablecoin type, we adopt an agent-based modeling approach and simulate each category. By varying different parameters and conditions in the simulations, we aim to identify under what circumstances each type of stablecoin is stable and not stable.

The study highlights potential risks associated with stablecoins with endogenous and centrally managed collateral, as they may be at a higher risk of crashing after a demand shock. A sudden surge in demand can trigger a death spiral, as seen in the TerraUSD crash. Moreover, our study shows that central collateral management may pose a greater risk of a bank run than decentralized collateral management because users may withdraw their funds en masse, leading to a collapse in the stablecoins’ value. Therefore, on one hand, stablecoins with exogenous and decentrally managed collateral, like Dai, may be considered the safest choice. On the other hand, stablecoins with endogenous and decentrally managed collateral, such as sUSD, provide greater autonomy and independence. However, these two designs also come at a cost: without additional measures, Dai and sUSD are more expensive to create and experience stronger fluctuations around their peg than others. Ultimately, we conclude that the choice between different types of stablecoins depends on individual risk preferences and specific use cases and that a trilemma exists between stability, independence, and costs.

Our study emphasizes the importance of carefully evaluating the source of a stablecoin’s collateral and its collateral management mechanism to ensure stability and minimize risks. Additionally, policymakers should exercise caution when dealing with stablecoins, such as TerraUSD, that have endogenous and \textit{centrally} managed collateral. To protect users and mitigate systemic risks in the financial system, it is crucial for policymakers to take steps to ensure the stability of these stablecoins through appropriate regulations. However, stablecoins that use a combination of endogenous and \textit{decentrally} managed collateral, such as sUSD, have proven stable and reliable. Therefore, regulators should carefully consider both dimensions when crafting policies rather than imposing overly restrictive measures on all stablecoins with endogenous collateral.

The paper is structured as follows. Section \ref{sec:2} discusses the literature on stablecoin categorization and presents our new 2 × 2 matrix. Section \ref{sec:3} explains the agent-based model and the simulation. Section \ref{sec:4} presents and discusses the results of the simulations. Finally, section \ref{sec:5} concludes.

\section{Stablecoin categorization}
\label{sec:2}
\subsection {State of the literature}
The most common categorization of stablecoins is based on the type of collateral. This classification typically yields three distinct types of stablecoins: fiat-collateralized, cryptocurrency-collateralized, and non-collateralized (or algorithmic) \cite{bank_for_international_settlements_investigating_2019, berentsen2019stablecoins,blockchain_state_2018,clark_demystifying_2020, grobys_stability_2021, kahya_reducing_2021,samman_state_2019,zhang2019regulation}. Aloui et al. (2021)\cite{aloui_are_2021} propose the concept of commodity-based stablecoins, such as those backed by gold, so commodity-collateralized constitutes a fourth type\cite{yadav_design_2021}. However, the literature utilizing this type of categorization often emphasizes the aspects of trust and decentralization instead of the stability of stablecoins.

Another approach categorizes stablecoins as either custodial or non-custodial \cite{zhao_understand_2021}. Custodial stablecoins depend on an issuer to hold reserves off-chain, whereas non-custodial stablecoins leverage smart contracts to establish a risk transfer market on the blockchain.

Bullmann et al. (2019)\cite{bullmann2019search} classify stablecoins with three dimensions: the presence or absence of an issuer responsible for fulfilling associated claims, the degree of decentralization of responsibilities in the stablecoin initiative, and the underlying value supporting stability of the stablecoin (source of value). Their approach allows us to classify any token within a 3 × 3 cube and has been utilized in other literature as well.\cite{kolodziejczyk_stablecoinstable_2020}. While the dimension “source of value” has a strong economic impact on price stability, the remaining two dimensions are less crucial in terms of an economic categorization.

Moin et al. (2020)\cite{moin_sok_2020} identify four dimensions of stablecoins without providing a classification
scheme as such. The dimensions are type of collateral, stability mechanism (e.g., reserve or dual coin), method for determining the price of the pegged asset (e.g., oracle, voting, trading), and peg used (e.g., pegged to fiat, a commodity, or a financial asset).

Finally, Klages-Mundt et al. (2020)\cite{klages-mundt_stablecoins_2020} and Klages-Mundt and Minca (2022)\cite{klages-mundt_while_2022} utilize the dimensions of custodial/non-custodial stablecoins and source of value. The source of value dimension is essential when examining stablecoin stability from an economic perspective, as exogenous vs. endogenous backing can have vastly different consequences, as we shall see. 

\subsection {New 2 × 2 matrix}
As evident from the literature, a generalized classification of stablecoins, such as the 3 × 3 cube by Bullmann\cite{bullmann2019search}, with an economic perspective has yet to be established. We fill this gap by introducing a two-dimensional categorization (Table \ref{tab:Table 1}) with a strong focus on economic principles. As the collateral drives the value of a stablecoin, we focus on the collateral in defining the dimensions collateral \textit{source} and collateral \textit{management}.

\vspace{\baselineskip}
\begin{table}[h]
\tiny
\centering
\resizebox{50em}{!}{%
\small
\begin{tabular}{lll}
\hline
& \multicolumn{2}{c}{Collateral management} \\
Collateral source & Central & Decentral \\ \hline
Exogenous & USDT\cite{TETHER}, USDC\cite{USDC} & Dai\cite{DAI} \\
Endogenous & TerraUSD\cite{kereiakes_terra_2019} & sUSD\cite{Synthetix} \\ \hline
\end{tabular}%
}
\caption{2 × 2 matrix for categorizing stablecoins with examples of each type.}
\label{tab:Table 1}
\end{table}

The first dimension is collateral source, derived from the “source of value” already used by Bullmann and Klages-Mundt\cite{Thisdimension}, which can be \textit{exogenous}, such as fiat and gold held in reserve, or \textit{endogenous}, in other words, a crypto asset of the same ecosystem\cite{erklärung}. The second dimension is collateral management, which can be \textit{central} entities and mechanisms managing pooled collateral or individuals managing their collateral \textit{decentrally}. This dimension has been discussed by e.g., Berentsen and Schär \cite{berentsen2019stablecoins} but not brought into a 2 × 2 matrix or 3 × 3 cube. Management differs from the custodial/non-custodial dimension as it centers on the issuer of stablecoins, i.e. who owns and manages the collateral, rather than the mode of issuance, whether through a central entity or a smart contract. These two dimensions are complementary and our 2 × 2 matrix could be expanded to a 3 × 3 cube by including the custodial /non-custodial dimension. For the purpose of this paper, however, we will stick with the 2 × 2 matrix. The matrix is applicable to any currency, on or off-chain. Appendix A shows the distinctions between the two stablecoin dimensions.

\section{Agent-based model and simulation}
\label{sec:3}
\subsection {Rationale for simulation}
Simulations can be a valuable tool to enhance our comprehension of the mechanisms behind the four types of stablecoins and to examine their stability conditions. However, it is difficult to predict the exact trajectory of stablecoin values, such as price and demand, using simulations. Market prices and demands are unpredictable due to chaos, external events, and individual decision-making. Furthermore, even if users’ decision-making of one stablecoin were understood, there is no guarantee that users of other stablecoins would behave the same way.

Notwithstanding the unpredictability of stablecoin values, simulations can help identify how stablecoin prices react to exogenous shocks. For example, simulations can be used to test the stability of stablecoins under different conditions and identify factors that can cause instability. The ability to test and refine control mechanisms in a simulated environment can be beneficial in reducing instability in real-world use. Nevertheless, our results should be analyzed and interpreted cautiously, keeping in mind that the model was not intended to perfectly replicate any particular real-world stablecoin but rather to illustrate differences in stablecoin design and response to shocks.

An agent-based approach is a suitable simulation method for testing stablecoin stability because it can produce a wide variety of unpredictable behavior with detail. Agent-based simulations have been used in simulating currencies \cite{mainelli_economic_2019} and can be adapted to simulate stablecoins. While success in controlling stablecoin values in a simulation is no guarantee of success in reality, failures of control in simulation can provide insights into the effectiveness of control mechanisms and identify areas for improvement.

\subsection {Specification of the model}
In this Monte Carlo experiment, the agents are \textit{users} who seek the stablecoin for non-volatile blockchain-based payments, \textit{investors} who buy and sell the stablecoin to earn fees,\cite{Investors} and \textit{issuers} who issue the stablecoin in exchange for collateral. Issuers differ with respect to collateral management. Under central collateral management, the issuer issues the stablecoin against collateral provided by the investor and then centrally manages the stablecoin. Under decentral collateral management, a protocol enables the investor to independently issue the stablecoin. Under decentral management, each investor can have one or more individual central-debt positions; the protocol intervenes only if the position is at risk of becoming under-collateralized.

Initially, the model sets up the parameters used for the price calculation of the endogenous or exogenous collateral as well as the initial wallets of the agents. Each wallet includes fiat money (in USD), stablecoins (valued in USD), and collateral (also valued in USD). The investors’ wallets are initialized with higher amounts than users’ to ensure a realistic representation. The model also sets up the number of paths and time steps.

The stablecoin demand of the investors heavily depends on users’ demand, and the investors adjust their demand accordingly. User demand in time step $t$ is calculated as follows:
\begin{equation}
    D_t(user) = \left\{\begin{array}{lr}
        (a - b \cdot fees + d \cdot s_t + r_t \cdot m), & \text{if } o_t \geq 1\\
        (a - b \cdot fees + d \cdot s_t + r_t \cdot m) \cdot o_t^2, & \text{if } o_{crit} \leq o_t < 1\\
        0 & \text{if } o_t < o_{crit}
        \end{array}\right\}.
\end{equation}

Variables $a$, $b$, and $d$ represent parameters used to calibrate the model. Investors charge users fees for trading with the stablecoin. A dummy variable $s_t$ indicates whether demand includes an exogenous shock\cite{demandshock}. Variable $r$ is a parameter that controls randomness in the system, while $m$ represents the magnitude of randomness. The collateral level of the stablecoin (i.e., the value of collateral divided by number of stablecoins) is described by $o$. If collateral level $o_t$ hits a critical level ($o_{crit}$), demand is assumed to be zero. The simulation is carried out over $t$ time steps.

Investor overall demand equals users’ demand plus a liquidity margin ($l$). In time step $t$ it is calculated as follows:
\begin{equation}
    D_t(investor) = l + D_t(user).
\end{equation}

Investors then create or burn stablecoins depending on the difference between their own holdings and their demand for stablecoins ($D_t(investors)$). 

We simulated a sequence of time steps for each path and modeled three different scenarios:
\begin{itemize}
       \item[1] \textit{Baseline}: No shock in user demand.
       \item[2] \textit{Negative}: A significant negative demand shock at a specific time step.
       \item[3] \textit{Positive}: A significant positive demand shock at a specific time step.
\end{itemize}
At each time step, a ten-step sequence of events occurs. First, the model updates the price of the stablecoin. The stablecoin price $P(s)$ at time step $t$ is calculated using the simple formula
\begin{equation}
    P_t(s) = D_t(s)/S_t(s)
\end{equation}
where $D_t(s)$ is the total demand, constituting the aggregate of user and investor demand, and $S_t(s)$ denotes the supply of the stablecoin, which refers to the quantity of stablecoins currently in circulation.

Second, the model calculates the price of the collateral. The price calculation differs depending on source. For stablecoin types like USDT and Dai, the collateral source is exogenous, and price is calculated using geometric Brownian motion:
\begin{equation}
    P_t(C_e) = P_0(C_e) \cdot \exp((\mu - 0.5 \cdot \sigma^2) \cdot t + \sigma \cdot W_t).
\end{equation}

$P_0(C_e)$ is the initial price of the (exogenous) collateral, $\mu$ is the expected return, and $\sigma$ is the volatility of collateral. $W_t$ is the standard Brownian motion process.

For stablecoin types like TerraUSD and sUSD, the collateral source is endogenous. We calculate its fair price considering that the collateral is in the same ecosystem as the stablecoin and holders/stakers of it receive (a share of) the trading revenues:
\begin{equation}
    P_t(C_i) = e \cdot (((D_t(user) \cdot fees)/z)/S_t(C_i))/c.
\end{equation}

In (5), $e$ represents a parameter used to calibrate the model; users are charged $fees$ by investors for users trading with the stablecoin (see also Eq. 1); $z$ represents a perpetual interest rate, while $c$ represents the opportunity costs, i.e., the return of an alternative investment option; $S_t(C_i)$ is the circulating supply of the (endogenous) collateral.

Third, the model calculates the stablecoin demand of the users (Eq. 1) and, fourth, the investors (Eq. 2).

Fifth, the model calculates the staking\cite{staking} demand of the investor. When considering staking, investors compare the returns on staking with other investment options. If the returns on staking increase—which is determined by stablecoin demand and stablecoin fees—staking becomes more appealing, and vice versa. We obtain the staking demand by solving for the amount of staking that makes investors indifferent to staking or not:
\begin{equation}
    D_t(staking) = f + g \cdot(D_t(user) \cdot fees)/c).
\end{equation}

In Eq. 6, $f$ and $g$\cite{bcanbe} represent parameters used to calibrate the model. Because the investors of stablecoin types with decentral collateral management can stake their endogenous collateral to receive stablecoins, this step applies only to stablecoin types with decentral collateral management, such as Dai and sUSD, and not to types like USDT and TerraUSD.

Sixth, in the case of central collateral management, if investor demand increases, investors buy stablecoins from the issuer. If investor demand decreases, investors sell stablecoins to the issuer. Investors pay a transaction fee for buying and selling to issuers. In the case of decentral collateral management, if investor demand increases, investors use the issuer protocol to issue central-debt positions by depositing collateral in at least the desired amount of stablecoins times the collateral ratio, allowing investors to issue stablecoins directly themselves. If investor demand decreases, investors use the issuer protocol to resolve central-debt positions.

Seventh, investors trade stablecoins with users based on user demand. If user demand increases compared to the previous time step, users buy stablecoins from investors, and if user demand decreases, users sell stablecoins to investors. In both cases, users pay a transaction fee to investors.

Eighth, if staking demand increased, investors stake additional collateral to receive stablecoins. If staking demand decreased, investors resolve some central-debt positions. This step applies only to stablecoin types with decentral collateral management, such as Dai or sUSD, and not to stablecoin types like USDT or TerraUSD.

Ninth, after each time step, the model saves all calculated values and updates each agent’s wallet as well as the supply and price of each asset.

Lastly, the model calls a supply check mechanism, which we discuss in the following subsection.

The second, fifth, sixth and eighth steps demonstrate the differences between the two dimensions of our 2 × 2 matrix, collateral \textit{source} and collateral \textit{management}. These steps involve calculations or actions that depend on one or both dimensions. Appendix B illustrates the algorithm of the agent-based model step-by-step.

\subsection {Control mechanisms}
We incorporate three control mechanisms in the model. The first ensures that the supply of every relevant asset, including stablecoins as well as endogenous and exogenous collateral, equals the sum of the specific asset amounts in agents’ wallets. This mechanism ensures that trades and other asset-producing or asset-destroying activities are accurately reflected in the supply of the asset and used to calculate the asset’s price. 

The second control mechanism focuses on executing the model’s functions correctly. It includes verifying that prices remain non-negative and ensuring that demand remains within a reasonable range. For instance, the mechanism prevents negative prices from arising, which can occur in exceptional circumstances. Additionally, it establishes bounds on the range of demand to maintain the model’s realism and prevent unrealistic scenarios. If any essential value is unavailable or any irregularity is detected, the model generates an error message to alert the simulator that there might be an issue with the system’s functionality.

The third control mechanism involves a function that checks and verifies the issuance and resolution of each central-debt position at each time step. By incorporating these control mechanisms, the model safeguards accuracy, consistency, and realism while providing error notifications in case of any discrepancies or anomalies.

\section{Results and discussion}
\label{sec:4}
Our results show that all four stablecoin types remain stable: in the absence of any demand shocks, they keep their peg. However, the types react differently after a significant negative demand shock; some even collapse. Decentralized stablecoins appear more resilient to bank runs than centralized stablecoins. Stability, independence, and costs constitute a trilemma.

\subsection {Demand shock}
\begin{figure}[h]
\centering
\subfloat[exogenous/centrally managed collateral]{\includegraphics[width=0.5\textwidth]{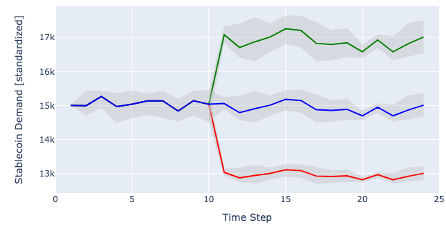}}
\subfloat[exogenous/decentrally managed collateral]{\includegraphics[width=0.5\textwidth]{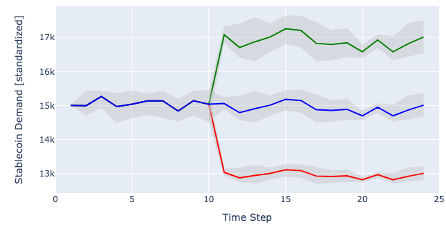}}\\
\subfloat[endogenous/centrally managed collateral]{\includegraphics[width=0.5\textwidth]{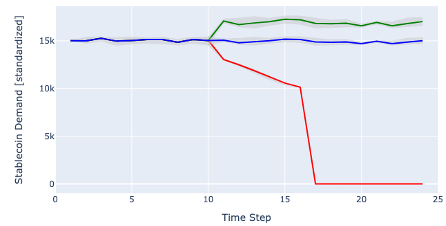}}
\subfloat[endogenous/decentrally managed collateral]{\includegraphics[width=0.5\textwidth]{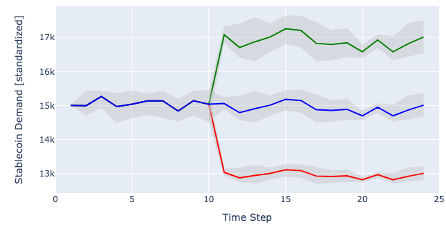}}\\
\caption{Evolution of stablecoin demand.} 
\label{fig:Figure 1}
\end{figure}

\begin{figure}[h]
\centering
\subfloat[exogenous/centrally managed collateral]{\includegraphics[width=0.5\textwidth]{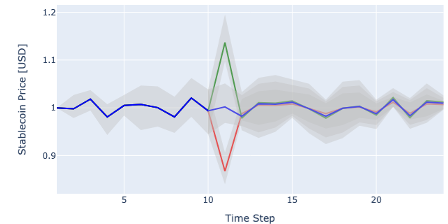}}
\subfloat[exogenous/decentrally managed collateral]{\includegraphics[width=0.5\textwidth]{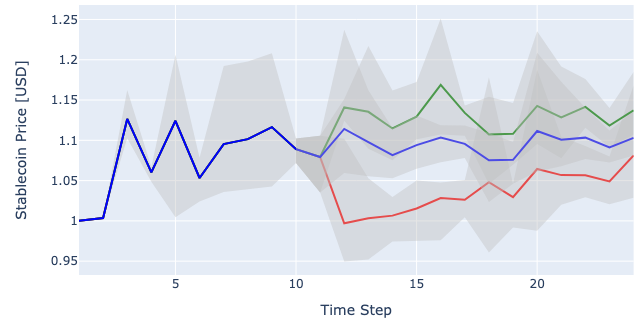}}\\
\subfloat[endogenous/centrally managed collateral]{\includegraphics[width=0.5\textwidth]{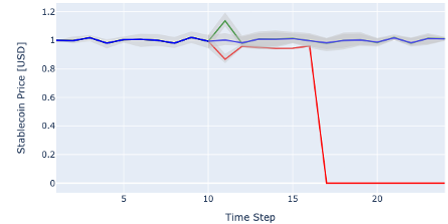}}
\subfloat[endogenous/decentrally managed collateral]{\includegraphics[width=0.5\textwidth]{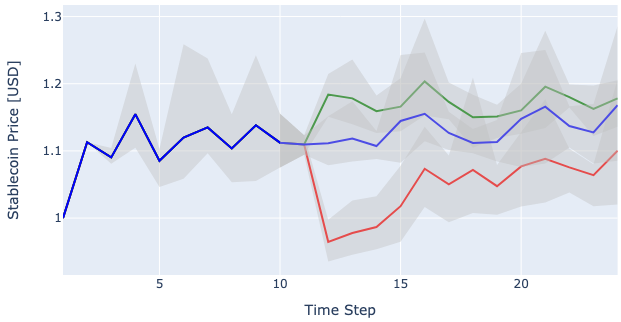}}\\
\caption{Evolution of stablecoin price in simulation.}
\label{fig:Figure 2}
\end{figure}

Figures \ref{fig:Figure 1} and \ref{fig:Figure 2} present the results of the simulation for the following scenarios: baseline model without a shock (blue line), positive demand shock (green line), and negative demand shock (red line). Results are shown for each dimension of the matrix: stablecoins with (a) exogenous and centrally managed collateral such as USDT, (b) exogenous and decentrally managed collateral such as Dai, (c) endogenous and centrally managed collateral such as TerraUSD, and (d) endogenous and decentrally managed collateral such as sUSD. Figure \ref{fig:Figure 1} depicts the evolution of stablecoin demand following a demand shock, while Figure \ref{fig:Figure 2} depicts the evolution of stablecoin price.

Figures \ref{fig:Figure 1}a and \ref{fig:Figure 1}b show that stablecoins with exogenous collateral, regardless of how the collateral is managed (centrally or decentrally), stabilize at a new equilibrium level of demand after experiencing a demand shock. The increase and decrease in demand remain consistent for positive and negative shocks. Figure \ref{fig:Figure 1}c and \ref{fig:Figure 1}d illustrate the divergent responses of stablecoins with endogenous collateral to negative demand shocks. Positive shocks elicit a similar effect for endogenous and exogenous collateral, but the reaction to negative demand shocks is markedly different. Stablecoins with endogenous and decentrally managed collateral cushion such shocks, while those with endogenous and centrally managed collateral are more vulnerable as demand fails to stabilize at a lower level and steadily decreases until it reaches zero.

The price evolution following a demand shock is shown in Figure \ref{fig:Figure 2}.

Figure \ref{fig:Figure 2}a shows that stablecoins with exogenous and centrally managed collateral typically experience an initial price increase in response to a positive demand shock and an initial price decrease following a negative shock. However, after a short time, the price recovers and eventually returns to (or close to) 1. This stablecoin type is resilient to demand shocks in the medium and long run. Figure \ref{fig:Figure 2}b demonstrates that stablecoins with exogenous and decentrally managed collateral have a less accurate peg and thus do not always perfectly track the price of 1 USD as they experience significant price fluctuations. However, they do not fully crash to 0, indicating they maintain at least some stability.

Figure \ref{fig:Figure 2}c shows a distinctive evolution for stablecoins with endogenous and centrally managed collateral. While the effect on price after a positive demand shock is initially similar to that of stablecoins with exogenous collateral, after a negative demand shock the price does not recover, and the stablecoin ultimately crashes. This crash is due to the ``death spiral."\cite{klages-mundt_while_2022,lyons_what_2023} For this stablecoin type, a decrease in demand leads to a reduction in future profits with the native coin, reducing the demand and value of the native coin. In addition, because the native coin backs the stablecoin, the collateral value of the stablecoin is also reduced. Once the collateral value hits a critical level, the demand for the stablecoin further decreases, exacerbating the decline in its price. Prior to reaching this critical value, the stablecoin's price does not immediately crash due to the presence of still-sufficient collateral backing. However, once the backing is no longer adequately high, the spiral becomes ``deadly" for the stablecoin, and the stablecoin crashes.

\begin{figure}[h]
\centering
\subfloat{\includegraphics[width=0.5\textwidth]{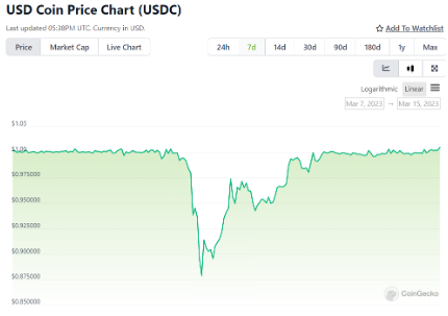}}
\subfloat{\includegraphics[width=0.5\textwidth]{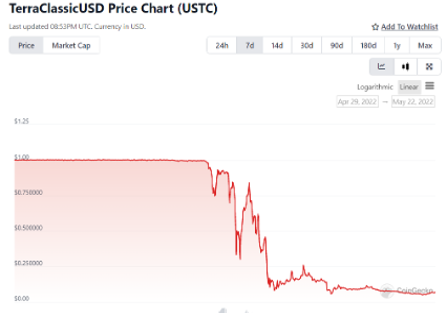}}\\
\caption{Evolution of stablecoin price in reality.\cite{Datafrom}}
\label{fig:Figure 3}
\end{figure}

The crash of TerraUSD in May 2022 is a real-life example of such a death spiral involving a stablecoin with endogenous and centrally managed collateral \cite{lee_dissecting_2023}. This example highlights the significance of prudence when dealing with such stablecoins, especially in uncertain environments. Conversely, the recent incident involving USDC, a stablecoin with exogenous collateral, also experienced a demand shock in March 2023. However, it managed to recuperate eventually, as predicted by our simulations. Figure \ref{fig:Figure 3} depicts the price evolution of these real-world scenarios.

Finally, Figure \ref{fig:Figure 2}d reveals a critical insight: stablecoins with endogenous and decentrally managed collateral exhibit similar behavior to those with exogenous and decentrally managed collateral. This finding emphasizes the significance of maintaining a high over-collateralization and liquidation ratio to safeguard against demand shocks and the resulting insecurity that could potentially trigger a death spiral. The liquidation mechanism ensures that positions that fall below a certain collateralization ratio are liquidated promptly to prevent under-collateralized debt positions. By implementing such risk-management practices, stablecoin issuers can enhance the stability and resilience of their systems.

It is worth noting, however, that stablecoins with endogenous and decentrally managed collateral still experience significant price fluctuations that exceed those of stablecoins with exogenous and decentrally managed collateral.\cite{Bycomparing} The more significant price fluctuations occur because the value of the collateral backing the stablecoin is directly tied to the value of the stablecoin itself, creating a feedback loop that can exacerbate price swings.

\subsection {Sensitivity analyses}
\label{subsec:sa}
To assess the robustness of our findings, we conduct sensitivity analyses on key factors including the magnitude of the demand shock, the volatility in demand, the fees, and the price of the (exogenous or the endogenous) collateral. These analyses allow us to explore the effects of deviations from our initial values in our model. The figures presenting the results for each stablecoin type and sensitivity analysis can be found in the Appendices C, D, E, and F.

We employ a systematic approach by testing five different sensitivity factors. Each factor involved multiplying the specific variable by 0.5, 0.75, 1, 1.25, and 1.5. This methodology allows us to observe the outcomes when deviating from our initial values, exploring both lower and higher ranges of up to 50\%. By employing this approach, we gain valuable insights into the impact of these deviations on our results, providing a comprehensive understanding of how variations in these factors influence stablecoin dynamics.

\begin{itemize}
       \item[1] \textit{Magnitude of demand shock}: For stablecoin types with exogenous/centrally managed collateral, exogenous/decentrally managed collateral, and endogenous/decentrally managed collateral, we observe consistent results across the five sensitivity factors. The primary difference we observed was in the initial deviations of the price, which were smaller or larger depending on the sensitivity factor. However, crucially, the price and demand of these types were able to recover over time. For stablecoin types with endogenous/centrally managed collateral, we observe similar results for positive demand shocks. However, for negative demand shocks, we observed that this type only experiences a crash if the magnitude of the demand shock is sufficiently large; i.e., for sensitivity factors of 0.5 and 0.75, this type was able to recover without crashing\cite{time}. Conversely, for sensitivity factors of 1.25 and 1.5, we observed a crash, which intuitively occurred more rapidly with a higher magnitude of the demand shock.
       \item[2] \textit{Volatility of demand}: Across all stablecoin types, we consistently observe the same outcomes across the five sensitivity factors. While there is increased volatility in both demand and price, the overall findings and conclusions remain unchanged.
       \item[3] \textit{Fees}: For stablecoin types with exogenous/centrally managed collateral, variations in fees did not have a significant effect on either the demand or the price. However, for stablecoin types with exogenous/decentrally managed collateral or endogenous/decentrally managed collateral, adjusting the fees led to changes in both the demand and subsequent price. Additionally, for stablecoin types with endogenous/centrally managed collateral, changes in fees can even result in a crash, independent of any external shocks.
       \item[4] \textit{Price of collateral}: For stablecoin types with exogenous/decentrally managed collateral, endogenous/decentrally managed collateral, and endogenous/centrally managed collateral, we observe a strong dependence of the stablecoin price on the price of the collateral. In our simulation, stablecoin types with exogenous/centrally managed collateral are directly backed by fiat money, which restricts our ability to analyze shocks in the collateral’s price. This limitation arises because fiat money serves as both the collateral and the reference currency in the model. However, if we were to adjust the model similarly to the other stablecoin types, we would expect to obtain comparable results.
\end{itemize}

These sensitivity analyses confirm the robustness of our results, emphasizing the instability of endogenous/centrally managed stablecoins and the significance of collateral and fees in stablecoin dynamics.

\subsection {Bank runs}
A bank run is where a large number of depositors withdraw their deposits from a bank because they fear that the bank may not have enough money to meet their withdrawal requests. Bank runs are often triggered by rumors or fears of a bank’s insolvency or bankruptcy. As depositors withdraw their money, other depositors start to panic and also withdraw their funds, leading to a decline in the bank’s reserves. In extreme cases, bank runs can result in the bank being unable to meet its obligations. During a bank run, the bank may be forced to sell its assets at a loss, which can lead to a chain reaction of other banks being affected and a disruption of the monetary system, causing a reduction in production \cite{diamond_bank_1983}.

In a centralized managed collateral system, a single entity—such as a central bank or a smart contract—controls the monetary system. As a result, depositors must rely on this entity’s ability to maintain the system’s stability and solvency. If the entity fails to do so, depositors may lose confidence in the system, triggering a bank run. In contrast, systems with decentral managed collaterals, such as Dai or Synthetix, do not have a single entity that controls the monetary system. Users rely instead on a distributed network of investors that maintain the system’s collateral individually. This decentralized structure means no single entity can manipulate or control the system, significantly reducing the risk of bank runs.

\subsection {Trilemma}
As the choice of stablecoin heavily depends on individual use cases and preferences, no single type of stablecoin can be considered the “best.” Instead, there are tradeoffs 
between (i) stability, (ii) independence, and (iii) costs. Stability refers to the ability of a system or currency to maintain a consistent value and avoid excessive fluctuations or crashing. Independence refers to the absence of a single issuer that centrally manages the collateral. This independence ensures that each individual stablecoin is always backed by collateral, thereby mitigating the risk of bank runs. Relevant costs are the sum of fees and opportunity costs due to over-collateralization.

The trilemma manifests itself in each of the four categories of stablecoins. These categories represent different approaches to achieving stability, independence, and costs.

\begin{itemize}
       \item[1] \textit{Exogenous/centrally managed collateral} (e.g., USDT): Stablecoins in this category do not require significant over-collateralization, making them cost-effective. However, their drawback is lack of independence. With central collateral management, investors face the risk of not being able to redeem the stablecoin for the collateral, resulting in bank-runs. In contrast, with decentralized collateral, users can always redeem the stablecoin by design (as long as their positions are not liquidated).
       \item[2] \textit{Exogenous/decentrally managed collateral} (e.g., Dai): Stablecoins in this category demonstrate stability in our simulation. However, they fluctuate more due to liquidations and higher required over-collateralization.
       \item[3] \textit{Endogenous/centrally managed collateral} (e.g., TerraUSD): Stablecoins in this category fail to prevent price crashes under many circumstances as seen in the sensitivity analysis, rendering them unreliable options. Their relative costs gains become inconsequential in light of this issue. Additionally, the independence concern raised for Tether-like stablecoins applies here, as this category relies on centralized collateral management.
       \item[4] \textit{Endogenous/decentrally managed collateral} (e.g., sUSD): Stablecoins in this category demonstrate stability in our simulation despite their endogenous backing, such as sUSD backed by the blockchain’s native token, SNX. However, achieving this stability comes at a higher cost, as these stablecoins need to be significantly over-collateralized. Like Dai-like stablecoins, they offer more independence but may also experience price fluctuations.
\end{itemize}

Overall, stablecoins with exogenous collateral like Dai or USDT are expected to perform well in terms of stability. However, they remain dependent on their collateral assets and lack the ability to react or counteract if the external collateral fails. On the other hand, sUSD-like stablecoins provide stability and independence through decentralized management and liquidations. However, the trilemma still exists due to the concerns, discussed earlier, about endogenous collateral and the less efficient issuance of sUSD, which requires over-collateralization at a rate of 400\%, compared to approximately 150\% for Dai and 100\% for Tether. 

Based on their characteristics, we can draw some general conclusions about the stability of each stablecoin type.\cite{Theexception} Dai offers high security against demand shocks and bank runs. However, without additional mechanisms, it is susceptible to significant price fluctuations, which may deter some users. USDT, although more stable, is more prone to bank runs due to centralized collateral management. While sUSD benefits from complete independence and decentralization, its endogenous collateral can lead to even larger price adjustments than Dai. These adjustments occur because the value of the collateral backing sUSD directly influences its value, creating a feedback loop that can amplify price swings. Despite this potential for increased volatility, sUSD’s decentralized and independent nature has made it popular among users valuing trustless and permissionless stablecoins.

Ultimately, the choice of stablecoin depends on individual preferences and priorities regarding stability, independence, and costs.

\subsection {Policy implications}
The emergence of stablecoins with centralized collateral management has led to concerns about the potential risks of bank runs. To mitigate these risks, policymakers should consider implementing regulations to ensure that such stablecoins maintain high levels of liquidity and sufficient collateral reserves. Stablecoins with an endogenous collateral source, such as Synthetix’s sUSD, should be subject to higher collateral requirements, such as over-collateralization (and associated liquidity ratio), to ensure stability.

Recently, a draft of a stablecoin bill\cite{kaloudis_draft_2023,us_house_financial_services_committee_stabelcoin_2023} argued that stablecoins with endogenous collateral sources might not be suitable for use. We suggest regulators should consider not only the collateral source but also collateral management, as stablecoins with endogenous and decentrally managed collateral have demonstrated their functionality thus far (e.g., sUSD). While 100\% stability cannot be guaranteed for any stablecoin, policymakers must remain vigilant and proactive in regulating stablecoins to ensure their safe and stable operation within the financial system. By regulating stablecoins, policymakers can mitigate the risks associated with these digital assets and promote a more stable and secure financial ecosystem.

\section{Conclusion}
\label{sec:5}
This paper presented a new approach to categorizing stablecoins based on an economic perspective and analyzed their stability under different scenarios. We identified a trilemma of stability, independence, and costs. The findings suggest that managing stablecoin collateral is critical in determining stability. We observed that stablecoins with endogenous collateral are more prone to instability and that centralized collateral management increases the risk of a bank run. Thus, policymakers must exercise caution with stablecoins with endogenous collateral and centralized collateral management, such as TerraUSD.

Future research should examine how different types of collateral, such as fiat money or cryptocurrencies, contribute to stablecoin stability. Exploring the impact of regulatory frameworks and interventions on stablecoin stability would provide valuable insights for policymakers. The effects of market liquidity on stability is another promising avenue. The dynamics of competition between various stablecoin types would also be an intriguing pursuit, yielding insights into market dynamics and the relative strengths and weaknesses of different stablecoin models. Moreover, the emergence of “flatcoins,” stablecoin variants designed to address inflation, warrants further investigation to determine the circumstances in which stablecoins or flatcoins offer greater suitability and advantages. The solid foundation established by the findings presented in this paper equips researchers and policymakers with a robust framework to assess and address stablecoin stability in these contexts, facilitating informed decision-making in the ever-evolving landscape of digital currencies.

\ledgernotes









\bibliographystyle{ledgerbib}
\bibliography{tempbib}

\begin{thebibliography}{10}
\bibitem{demandshock}
In our initial model, we incorporate a shock of 40\% at time step 10 of 24.
  However, in our sensitivity analysis (subsection \ref{subsec:sa}), we explore
  the results with both larger and smaller shocks to provide a comprehensive
  examination.
\bibitem{bcanbe}
``$g$" can be a positive (e.g., {sUSD}) or negative (e.g., Dai) number.
\bibitem{Bycomparing}
Comparing historical price data of Dai and sUSD, it is evident that the price
  volatility of sUSD has indeed been significantly greater than that of Dai.
\bibitem{time}
Through simulations with an increased number of time steps, we have verified
  that this type exhibits stability not only immediately following smaller
  magnitude shocks but also in the long run.
\bibitem{Theexception}
The exception is TerraUSD. Due to its centrally managed and endogenous
  collateral, this type may not offer significant advantages over the other
  types of stablecoins.
\end{thebibliography}


\newpage 	

\renewcommand{\thesection}{\Alph{section}}
\appendix
\setcounter{section}{0}
\makeatletter
\renewcommand{\p@subsection}{\thesection.}
\makeatother

\section{Distinctions among stablecoin dimensions}
\label{appendix:A}

\begin{table}[h]
\tiny
\centering
\resizebox{87em}{!}{%
\small
\begin{tabular}{lll}
\hline
Dimension  & Mechanism \\ \hline
Central collateral management  & Collateral is pooled together and the issuer functions as a central entity or is a smart contract responsible for managing the collateral. \\
Decentral collateral management  & “Everyone” has the ability to issue central-debt positions, which are subject to liquidation once a specific critical liquidation ratio is reached. \\
Exogenous collateral source & External to the blockchain system; the collateral has no interactions with the stablecoin other than backing the stablecoin. \\
Endogenous collateral source & Internal to the blockchain system; stablecoin value is derived from future trading fees or other financial remunerations. \\ \hline
\end{tabular}%
}
\label{tab:ap2}
\end{table}

\section{Agent-based model algorithm}
\label{appendix:B}

\begin{figure}[h]
  \centering
  \includegraphics[width=0.41\textwidth]{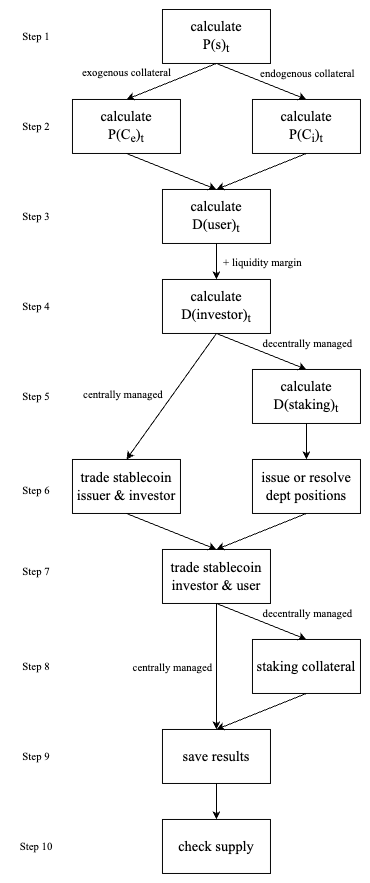}
  \label{fig:algo}
\end{figure}
\newpage

\section{Sensitivity analyses - Exogenous/centrally managed collateral}
\label{appendix:C}
\begin{figure}[h]
  \centering
  \includegraphics[width=1\textwidth]{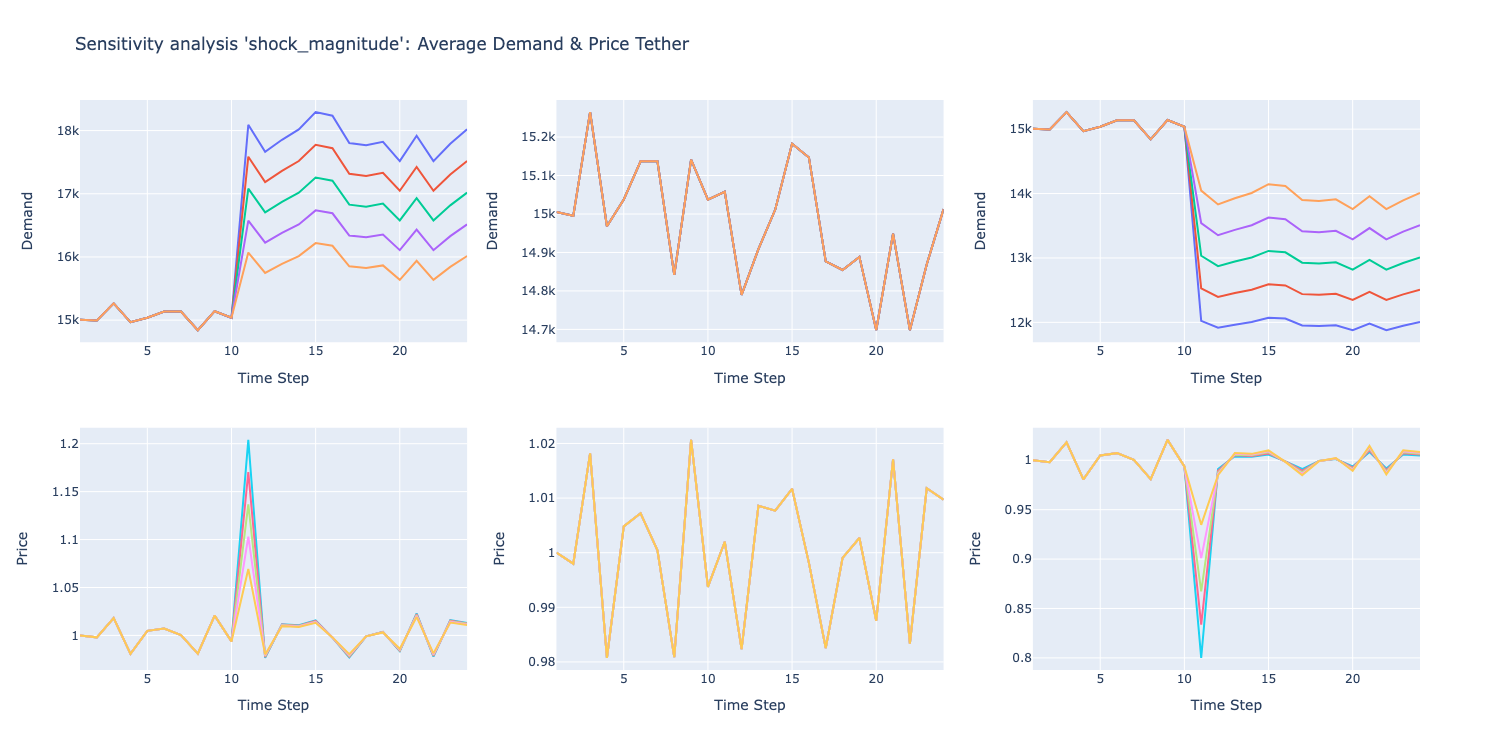}
\end{figure}
\begin{figure}[h]
  \centering
  \includegraphics[width=1\textwidth]{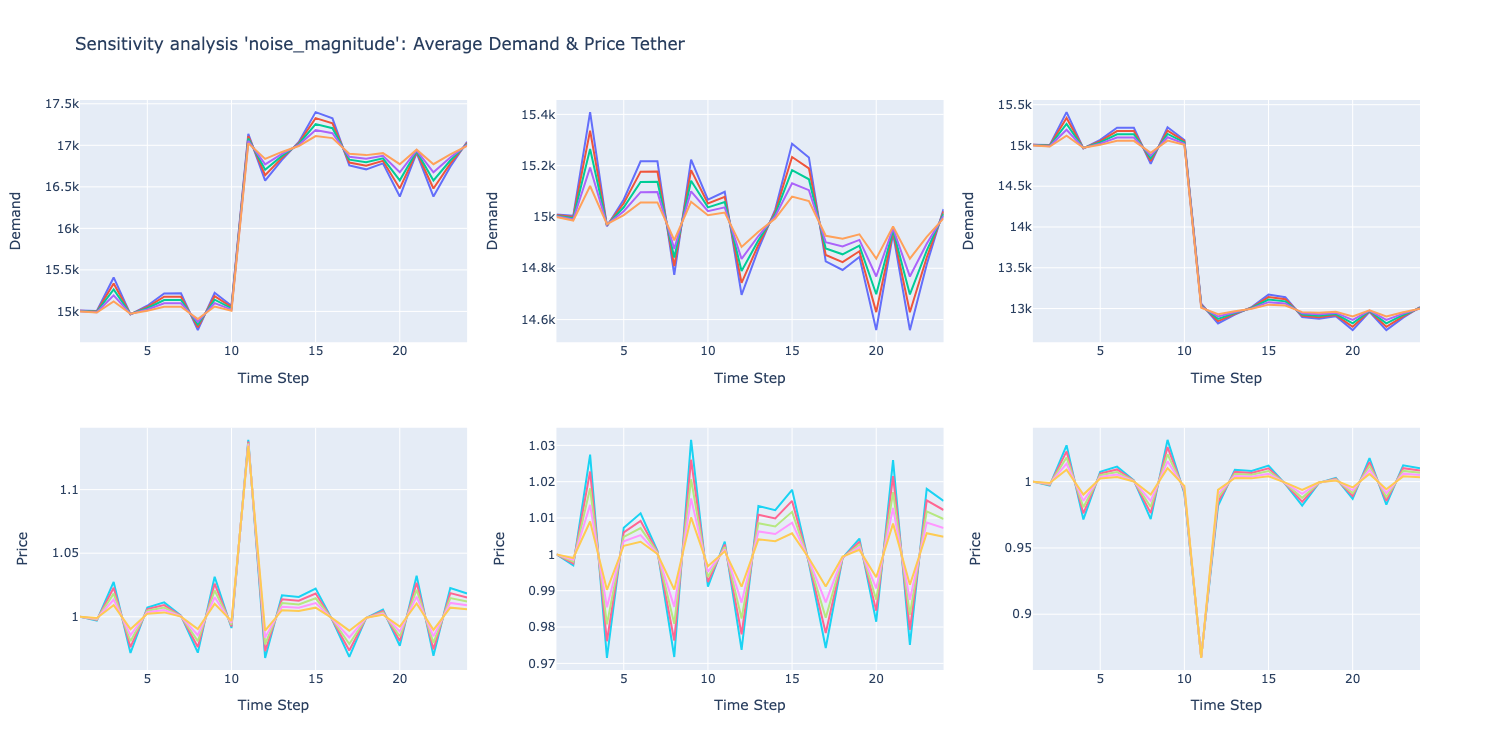}
\end{figure}
\newpage
\begin{figure}[h]
  \centering
  \includegraphics[width=1\textwidth]{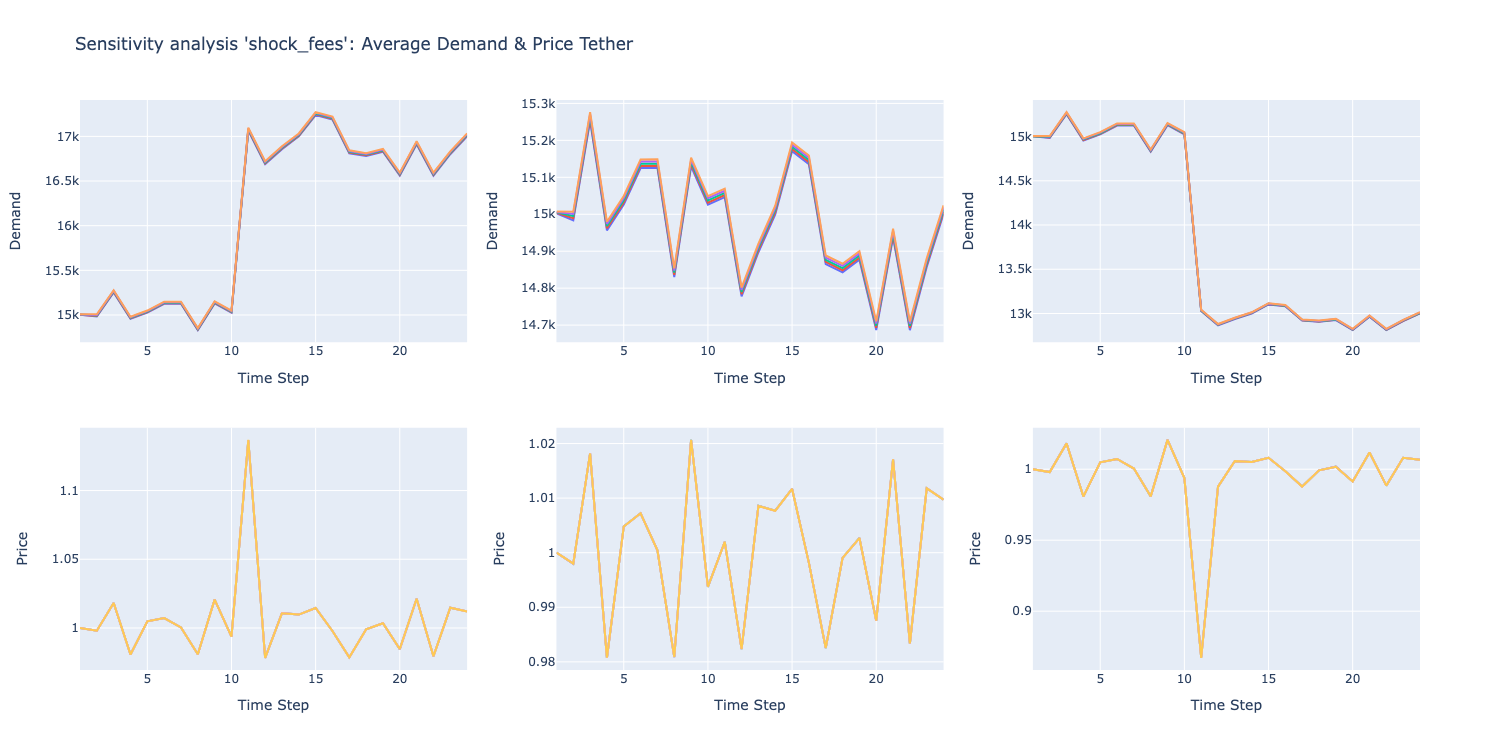}
\end{figure}
\begin{figure}[h]
  \centering
  \includegraphics[width=1\textwidth]{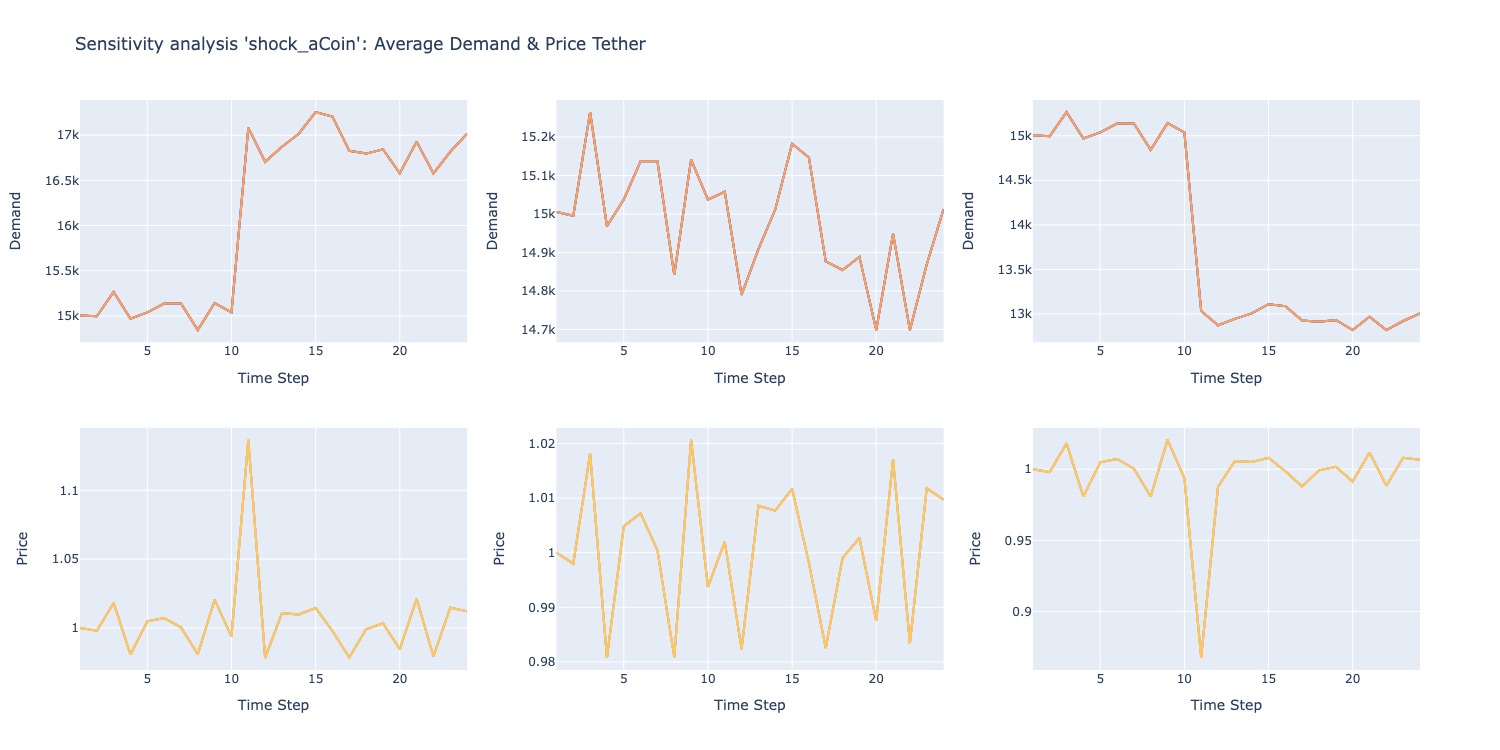}
\end{figure}
\newpage

\section{Sensitivity analyses - Exogenous/decentrally managed collateral}
\label{appendix:D}
\begin{figure}[h!]
  \centering
  \includegraphics[width=1\textwidth]{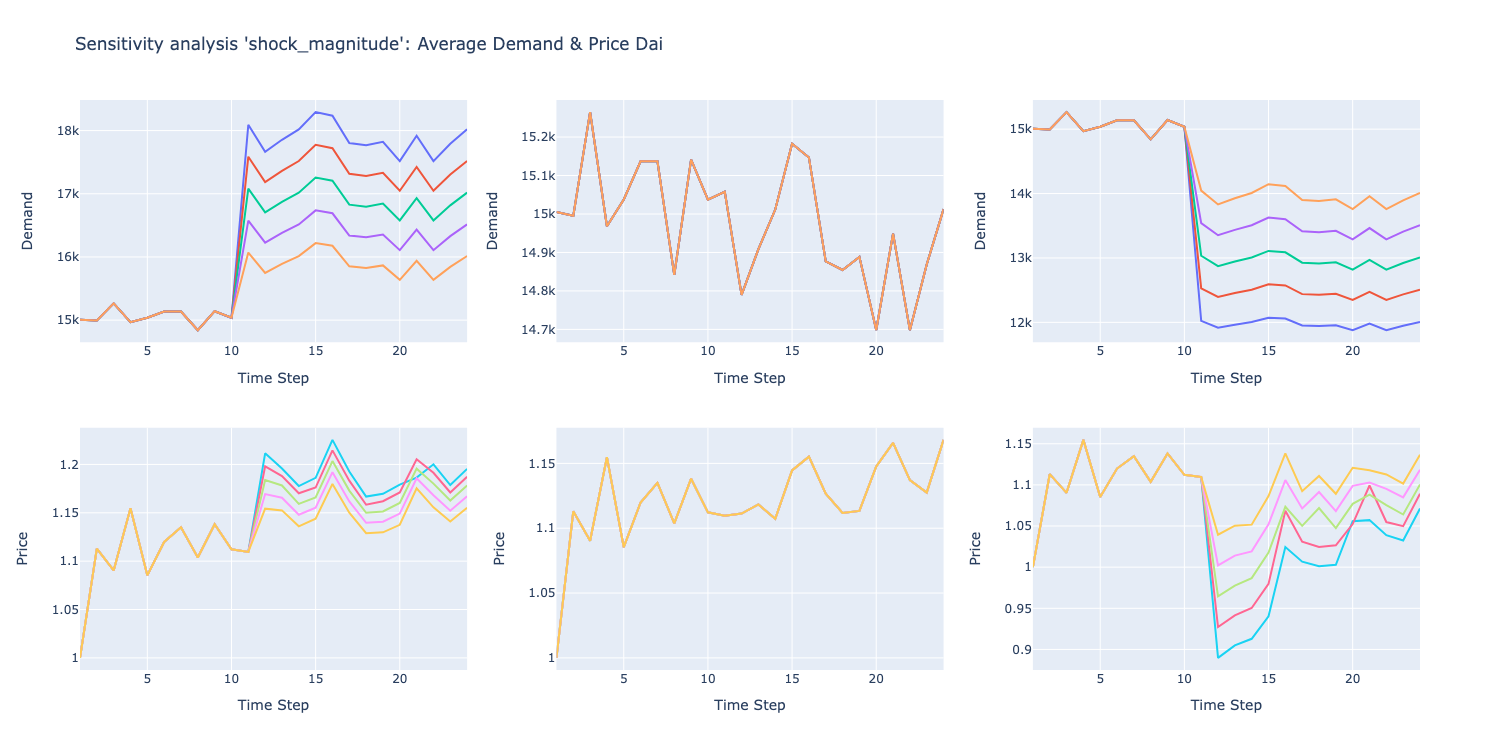}
\end{figure}
\begin{figure}[h!]
  \centering
  \includegraphics[width=1\textwidth]{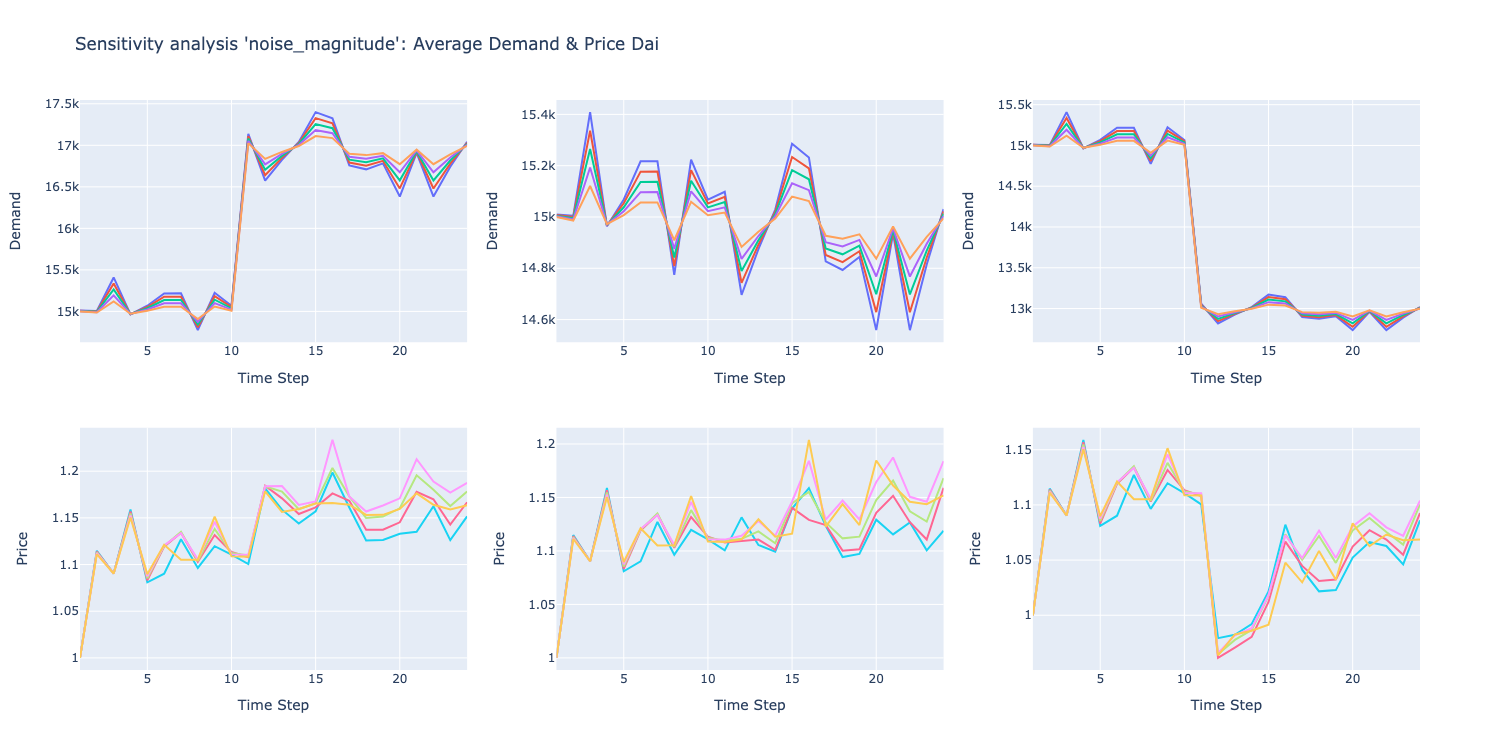}
\end{figure}
\newpage
\begin{figure}[h!]
  \centering
  \includegraphics[width=1\textwidth]{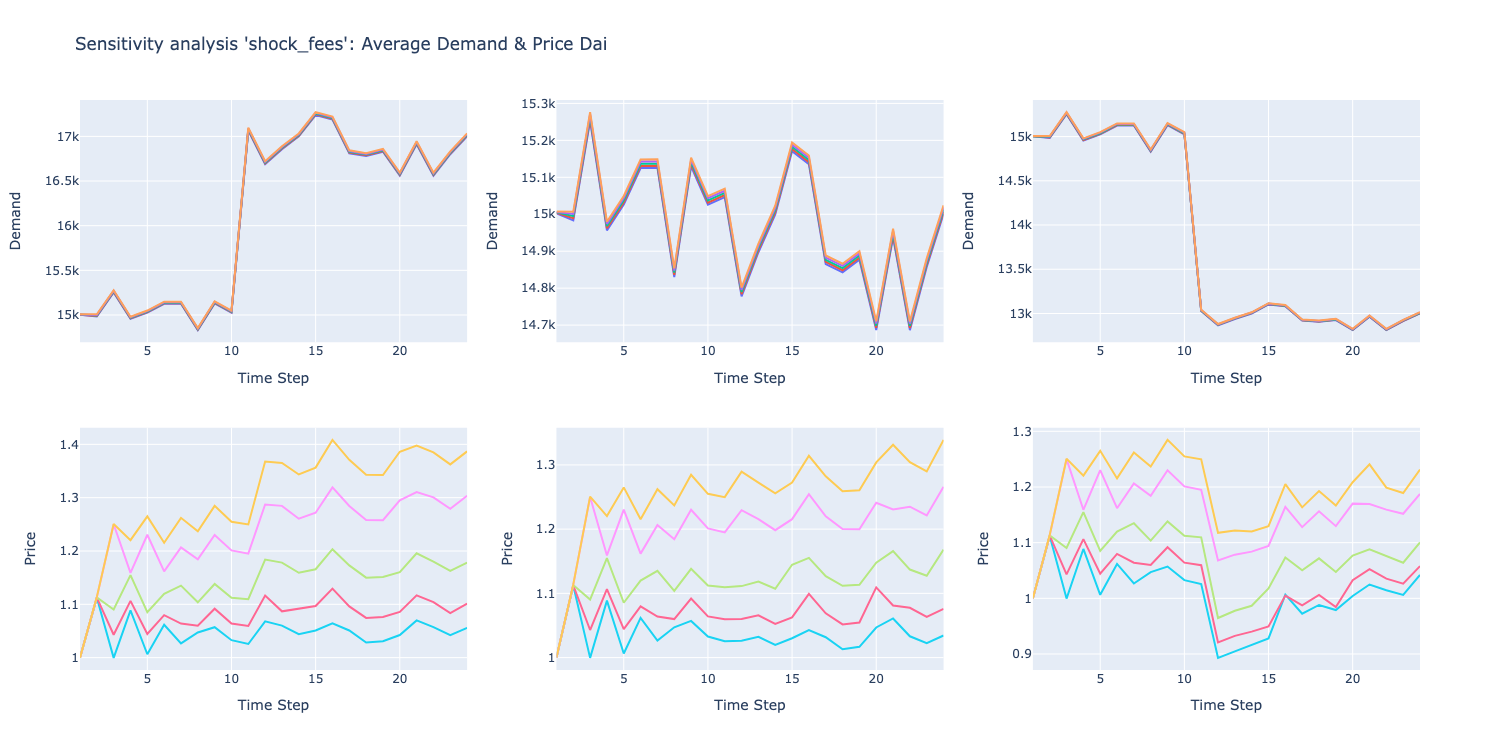}
\end{figure}
\begin{figure}[h]
  \centering
  \includegraphics[width=1\textwidth]{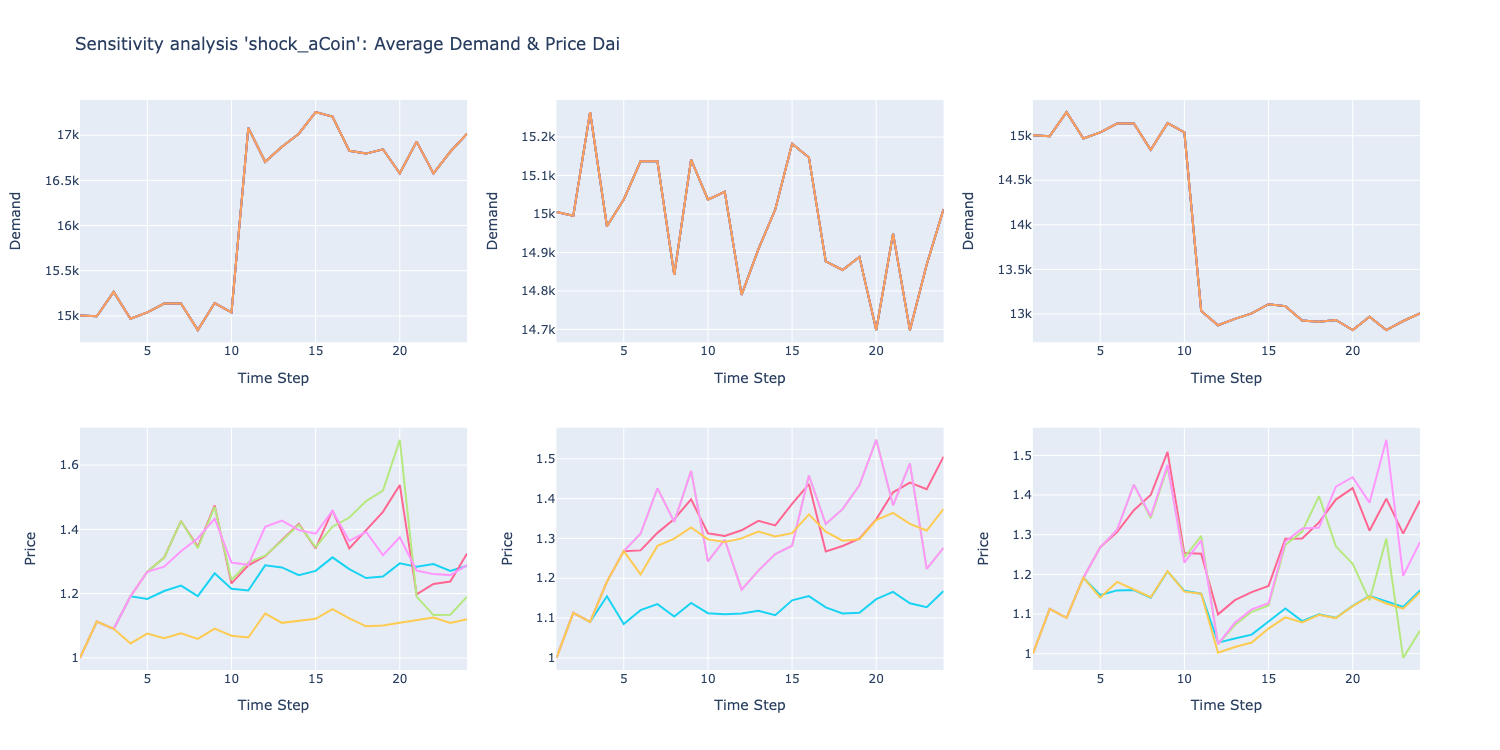}
\end{figure}
\newpage

\section{Sensitivity analyses - Endogenous/centrally managed collateral}
\label{appendix:E}
\begin{figure}[h]
  \centering
  \includegraphics[width=1\textwidth]{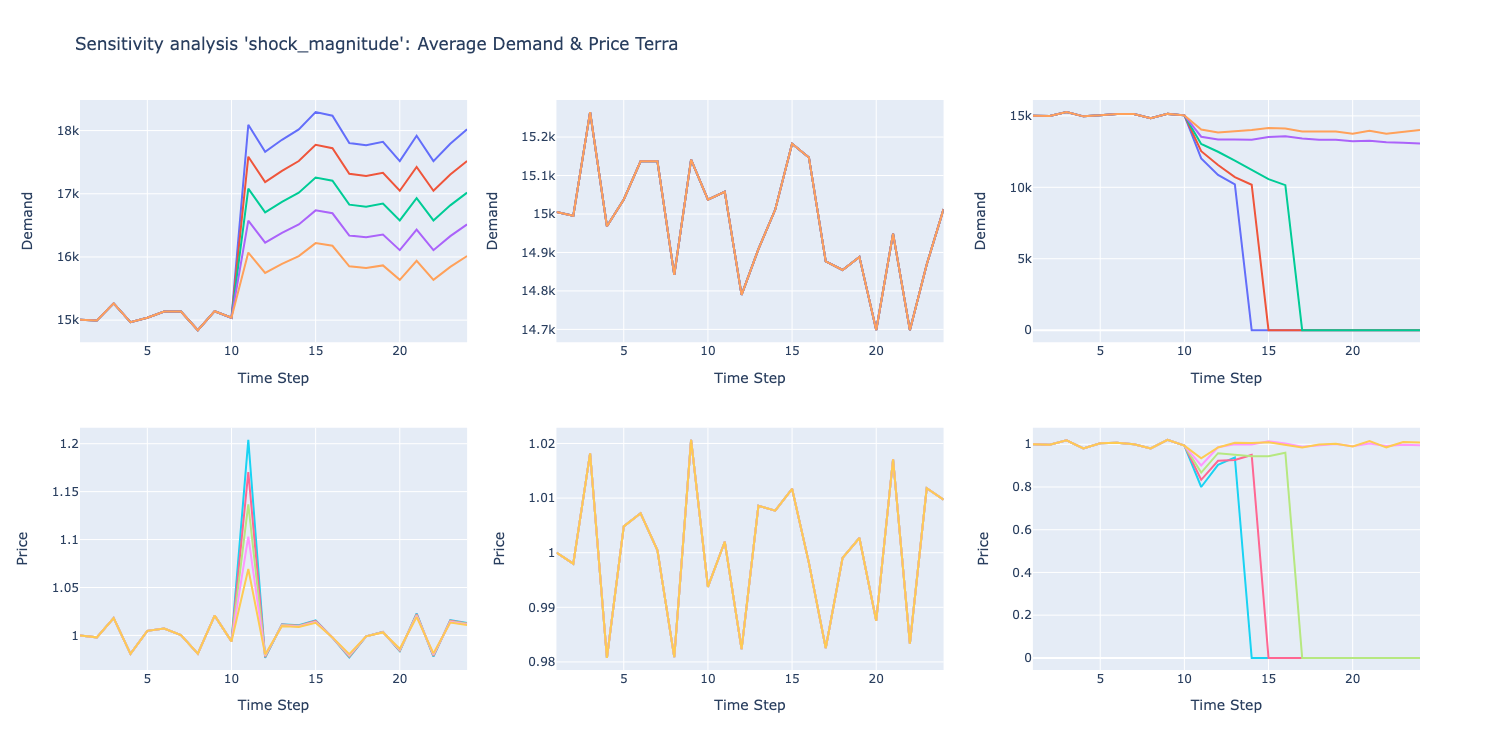}
\end{figure}
\begin{figure}[h]
  \centering
  \includegraphics[width=1\textwidth]{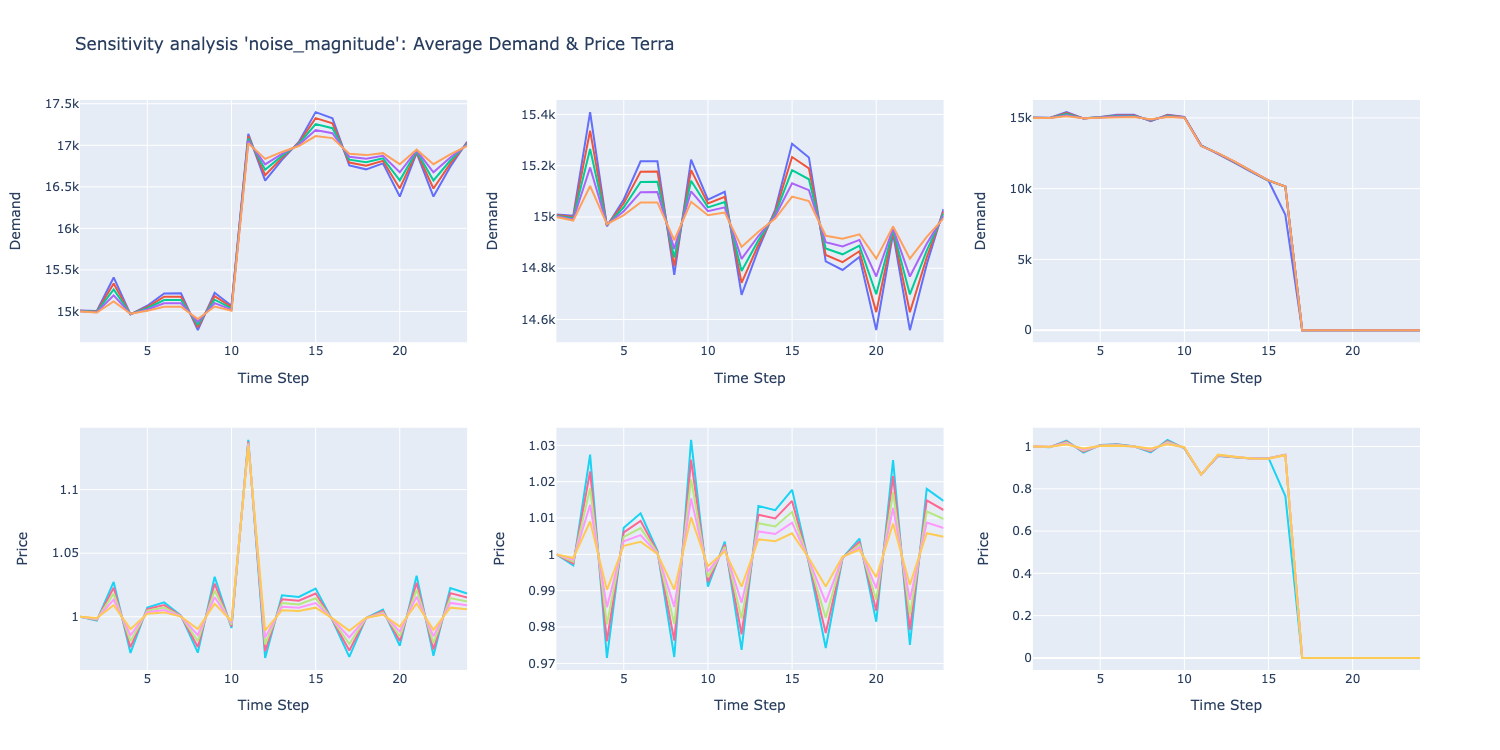}
\end{figure}
\newpage
\begin{figure}[h]
  \centering
  \includegraphics[width=1\textwidth]{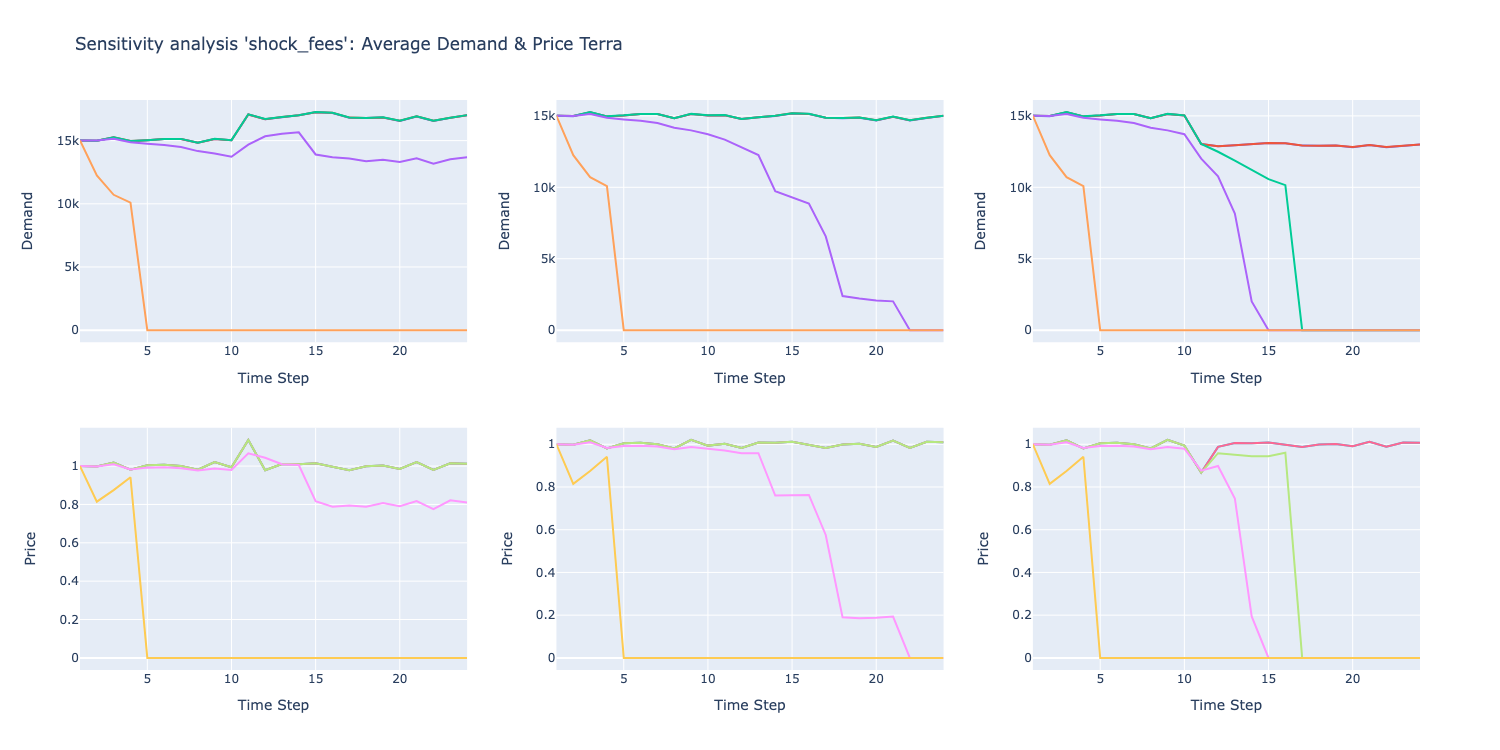}
\end{figure}
\begin{figure}[h]
  \centering
  \includegraphics[width=1\textwidth]{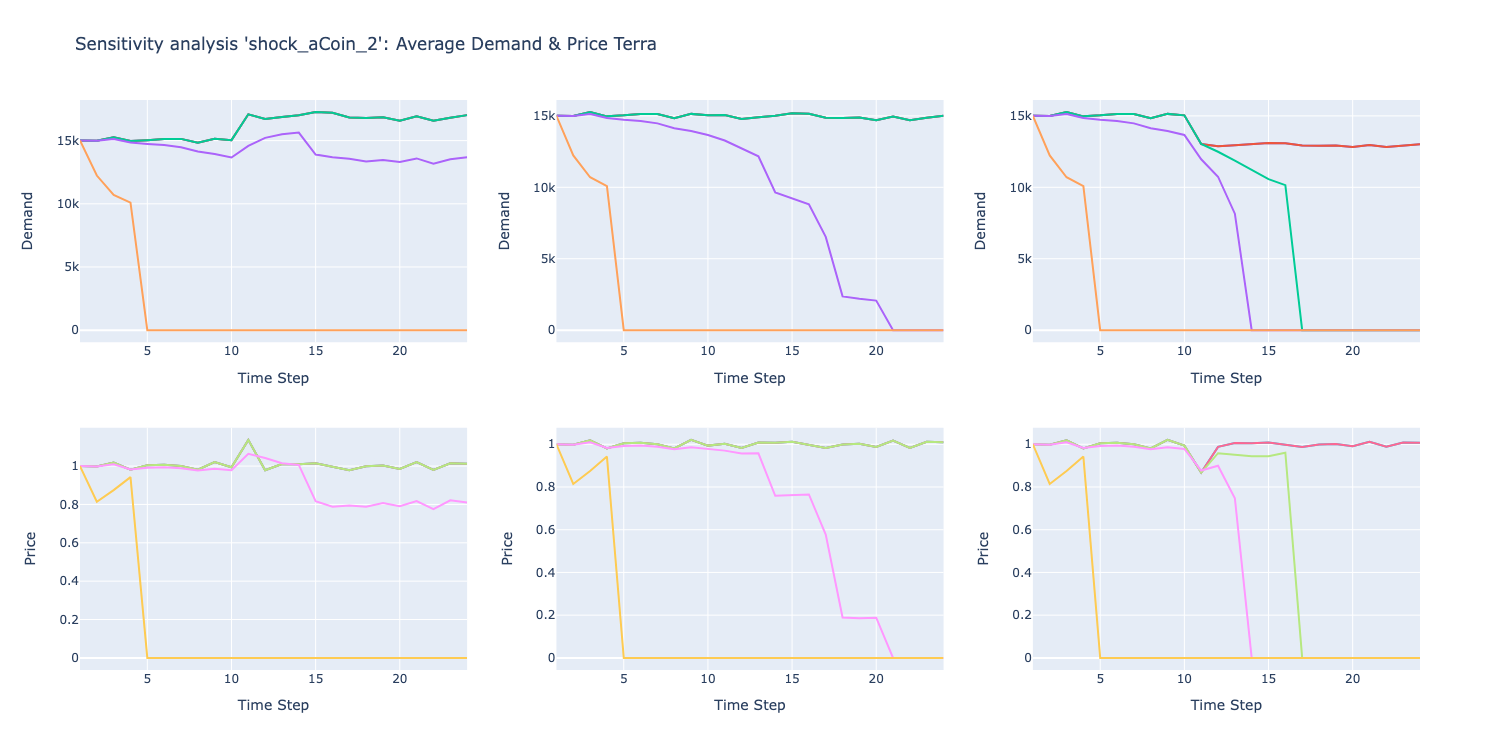}
\end{figure}
\newpage

\section{Sensitivity analyses - Endogenous/decentrally managed collateral}
\label{appendix:F}
\begin{figure}[h]
  \centering
  \includegraphics[width=1\textwidth]{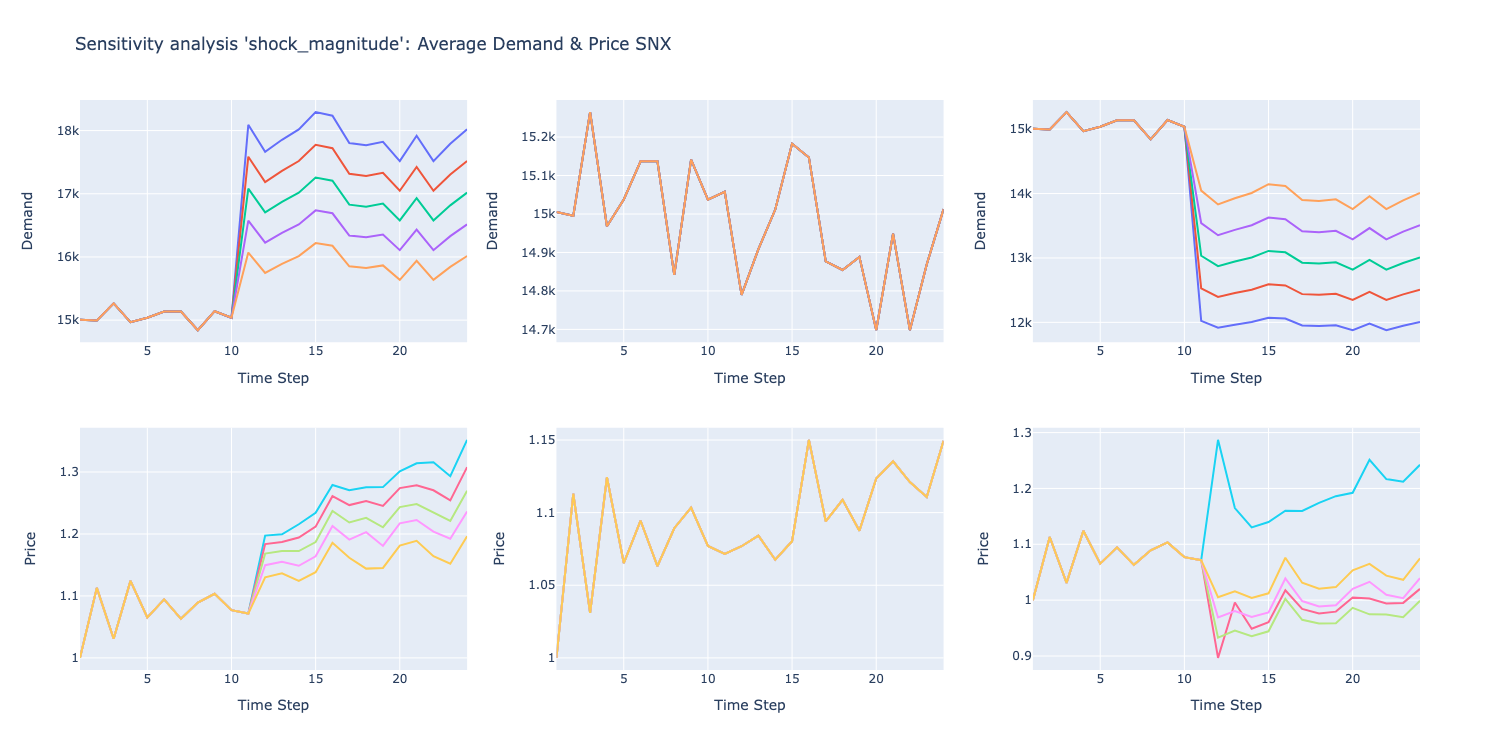}
\end{figure}
\begin{figure}[h]
  \centering
  \includegraphics[width=1\textwidth]{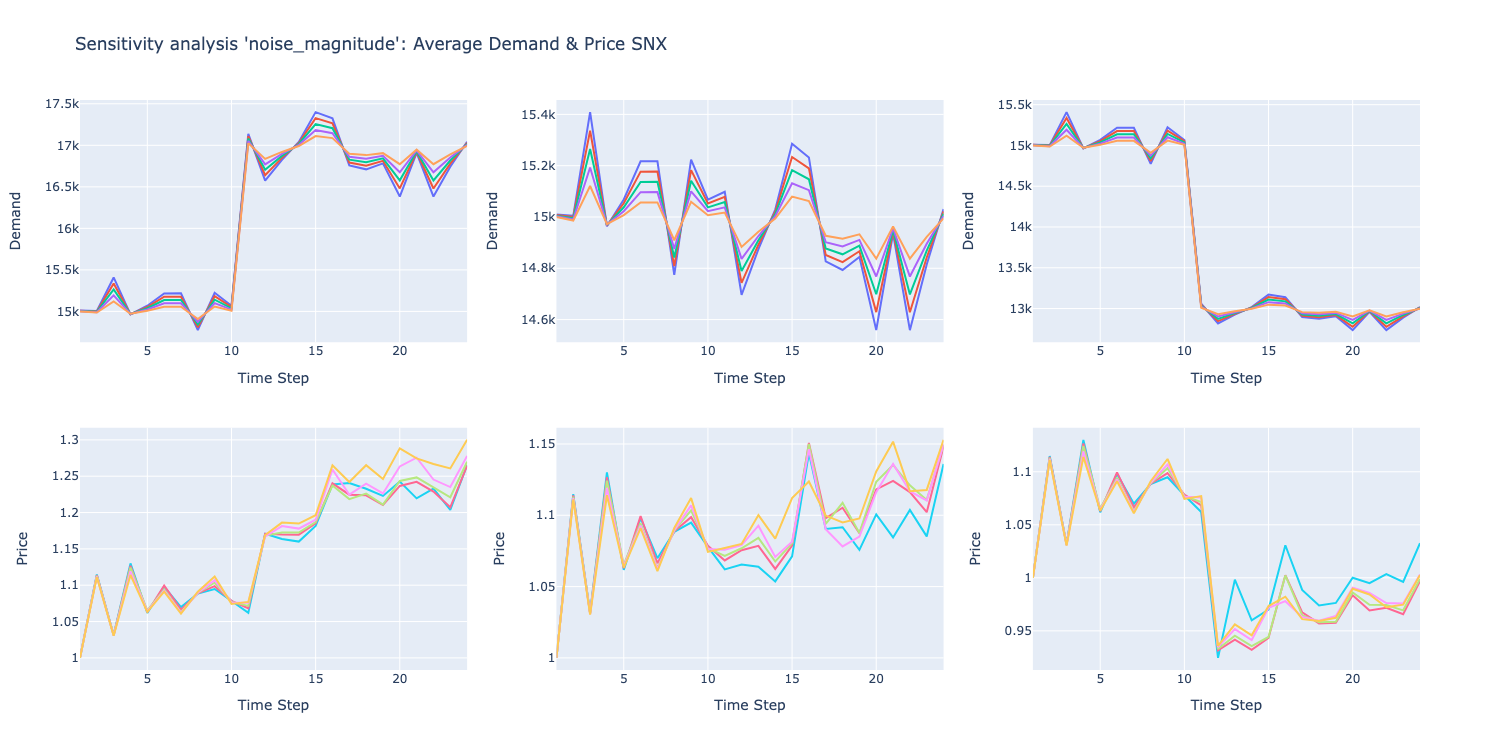}
\end{figure}
\newpage
\begin{figure}[h]
  \centering
  \includegraphics[width=1\textwidth]{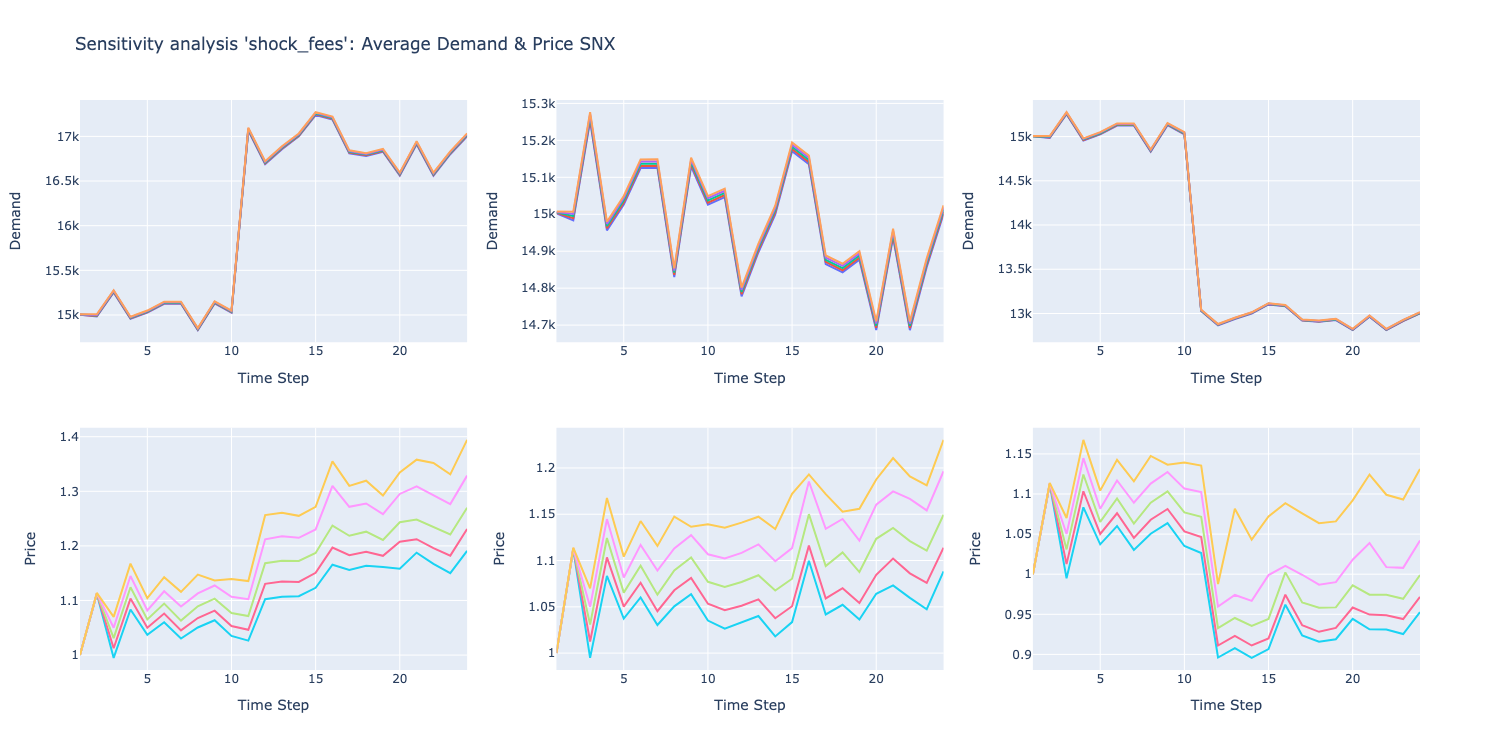}
\end{figure}
\begin{figure}[h]
  \centering
  \includegraphics[width=1\textwidth]{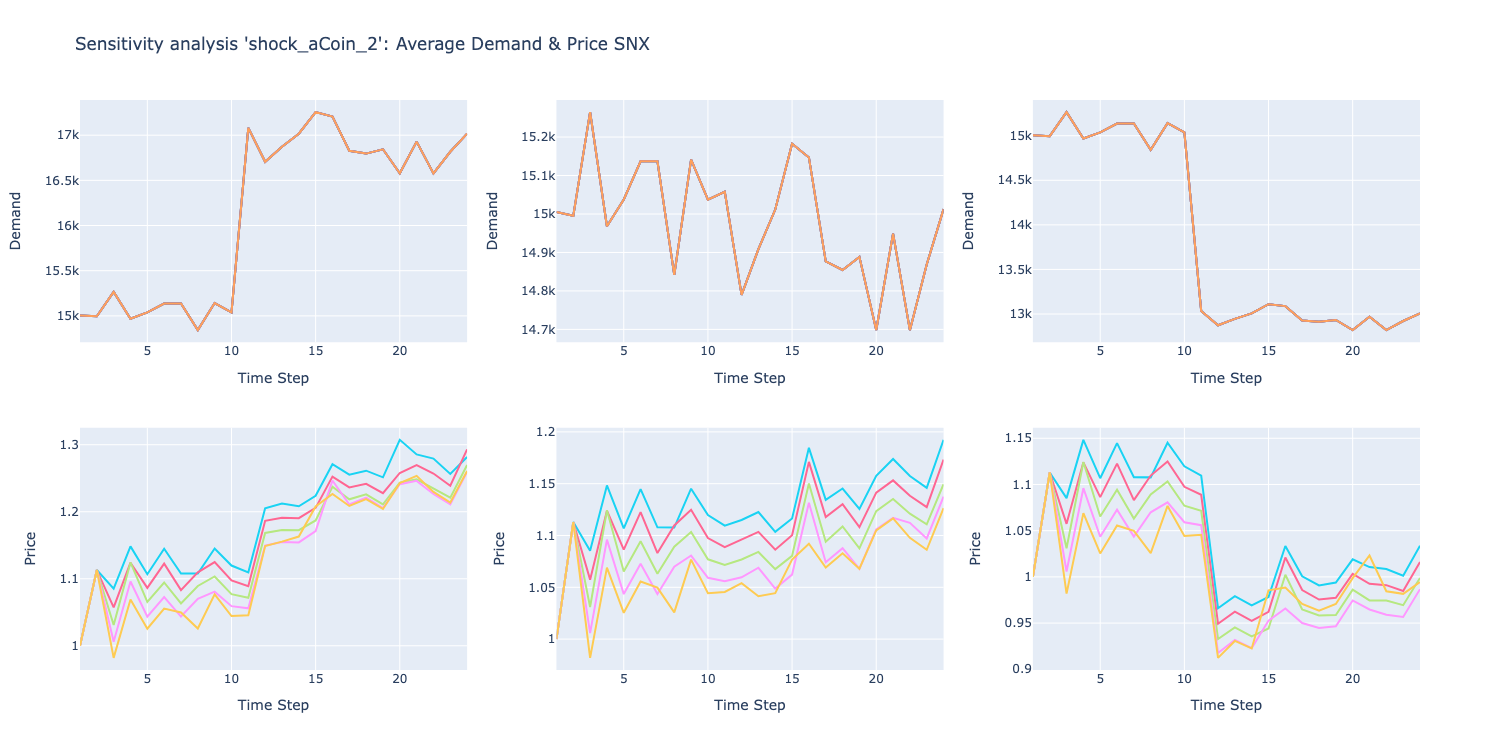}
\end{figure}






\end{document}